\documentclass[aps,prl,twocolumn,superscriptaddress,groupedaddress]{revtex4}  
\usepackage{graphicx}  
\usepackage{dcolumn}   
\usepackage{bm}        
\usepackage{amssymb}   
\usepackage{siunitx}
\usepackage[breaklinks=true,colorlinks,citecolor=blue,linkcolor=blue,urlcolor=blue]{hyperref}
\usepackage{multirow}
\usepackage{xcolor}
\usepackage{amsmath}
\usepackage{lipsum}

\usepackage{color} 
\begin{document}

\title{A Higher-Order Topological Insulator Phase in a Modulated Haldane Model}

\author{Baokai Wang}
\thanks{B. W. and X. Z. contributed equally to this work.}
\affiliation{Department of Physics, Northeastern University, Boston, Massachusetts 02115, USA}

\author{Xiaoting Zhou}
\thanks{B. W. and X. Z. contributed equally to this work.}
\affiliation{Department of Physics, Northeastern University, Boston, Massachusetts 02115, USA}

\author{Hsin Lin}
\affiliation{Institute of Physics, Academia Sinica, Taipei 11529, Taiwan}

\author{Arun Bansil}
\email{ar.bansil@northeastern.edu}
\affiliation{Department of Physics, Northeastern University, Boston, Massachusetts 02115, USA}


\begin{abstract}

We explore topological properties of a modulated Haldane model (MHM) in which the strength of the nearest-neighbor and next-nearest-neighbor terms is made unequal and the three-fold rotational symmetry $\mathcal{C}_3$ is broken by introducing a trimerization term ($|t_{1w(2w)}|< t_{1s(2s)}$) in the Hamiltonian. Using the parameter $\eta={t_{1w}}/{t_{1s}}= t_{2w}/t_{2s}$, we show that the MHM supports a transition from the quantum anomalous Hall insulator (QAHI) to a HOTI phase at $\eta=\pm 0.5$. The MHM also hosts a zero-energy corner mode on a nano-disk that can transition to a trivial insulator without gap-closing when the inversion symmetry is broken. The gap-closing critical states are found to be magnetic semimetals with a single Dirac node which, unlike the classic Haldane model, can move along the high-symmetry lines in the Brillouin zone. MHM offers a rich tapestry of HOTI and other topological and non-topological phases.

\end{abstract}

\maketitle

{\it Introduction.--}
The discovery of topological insulators (TIs)\cite{TI_3, TI_1, TI_2, TI_4} has triggered a rapidly developing  research field. The signature of a $d$-dimensional TI is the emergence of robust, symmetry-protected gapless states on its $(d-1)$-dimensional boundaries. Recently, topological crystalline insulators (TCIs) with higher-order bulk-boundary correspondence have been discovered, which are dubbed as higher-order topological insulators (HOTIs). A HOTI exhibits a gapped $(d-1)$-dimensional boundary and supports topologically protected states on a lower $(d-n)$-dimensional boundary where $n \geqslant 2$. 
For example, a three-dimensional (3D) HOTI exhibits 2D gapped surfaces and topologically-protected states on its one-dimensional (1D) hinges. Similarly, a two-dimensional (2D) HOTI hosts nontrivial states on the 0D corners of a nanodisk rather than the gapless propagating 1D edge states.
HOTIs have been proposed and realized in a variety of 3D and 2D materials\cite{hoti_bi, EZAWA_1, EZAWA_2, anomalous_hoti, soti_crystalline, multiple, strong_weak_t3, hoti_inversion, lattice_hoti, hoti_rotoinversion, cn_hoti, chiral_1, protected, zero, graphyne, antikekule, hoti_tci, majorana_hoti, symmetry_hierarchy, third, tci, fractional_1, graphdiyne, elastic, general, aah, cutting, hoti_sb, brick, engieering_hoti, squareroot, symmetryenforced, universal, bisthmus_hans, bi4br4, glidemirror}, as well as in photonic and phononic systems\cite{photonic_1, acoustic_1, coldatom_1, acoustic_nm, acoustic_nm_chiral, sound, sonic, holey_acoustic, advances_photonics, phononic_1, phononic_2, photonic_exp_1, photonic_exp_2, third_acoustic, acoustic_exp_1, acoustic_exp_2, photonic_kagome, anomalous_quadrupole, third_hierarchy}, quasicrystals\cite{quasicrystal, amorphous}, heterostructures\cite{heterostructure, tbg}, magnetic compounds\cite{magnetic, ferromagnetic, hinged_qshe}, topoelectrical circuits\cite{octupole, topoelectric}, interacting fermion and boson systems\cite{hoti_interacting_1, spinliquid, interaction, mott, ssh_interacting, luttinger, boson}, non-Hermitian\cite{nonhermitian_1, nonhermitian_2, nonhermitian_3, antiunitary, gainloss, nonhermitian_4, nonhermitian_sonic, nonhermitian_floquet} and other systems\cite{disordered, periodic_1, hoti_external, solitonic, light, hightemp, floquet_1, periodic_driving, floquet_hoti, singularity}.  

\begin{figure}
\includegraphics[width=\linewidth]{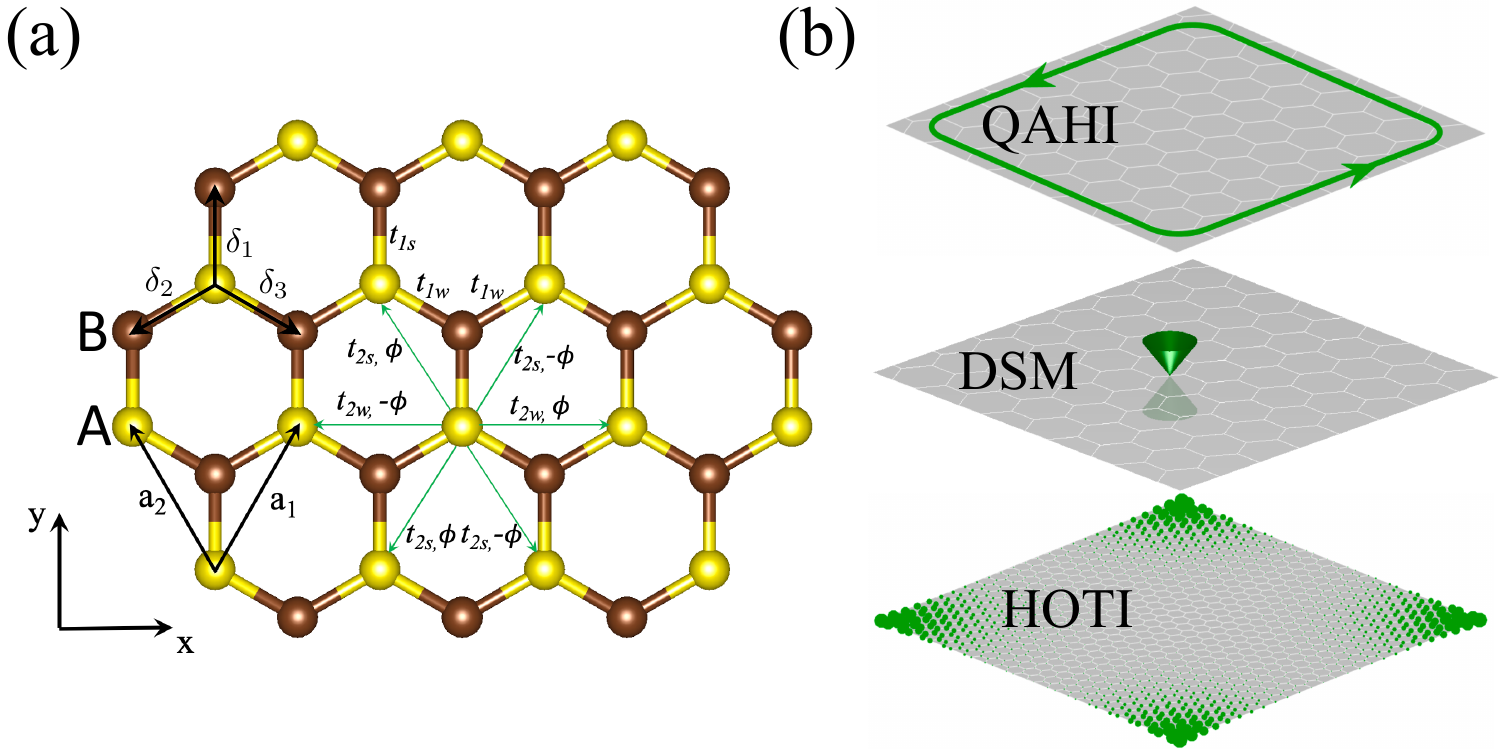}  
\caption{(a) The modulated Haldane model (MHM) on the honeycomb lattice with A (yellow) and B (brown) sublattices. The lattice translation vectors are $\mathbf{a}_{1,2}$. $t_{1s,1w}$ ($t_{2s,2w}$) denotes the strength of the NN (NNN) hopping, with $t_{1s}$ ($t_{2s}$) stronger than $t_{1w}$ ($t_{2w}$).  $\phi$ is the magnetic phase. (b) A schematic of the phase transitions realized in MHM, from a QAHI to a HOTI, via the critical semimetal state hosting a single Dirac cone.}
\label{fig1}
\end{figure}

Haldane model, which has become a classic model in the field of topological materials, realizes the quantum Hall effect on a honeycomb lattice without an external magnetic field\cite{Haldane}, also known as quantum anomalous Hall insulator (QAHI). The introduction of magnetic phases $\phi$ on the next-nearest-neighbor (NNN) hoppings breaks the time reversal (TR) symmetry $\mathcal{T}$ without net magnetic flux per plaquette.  Here, we introduce a modulated Haldane model in which the three-fold rotational symmetry $\mathcal{C}_{3}$ is broken via the trimerization in the hopping along the $\mathbf{\delta}_1$ direction as shown in Fig.~\ref{fig1}(a), where $|t_{1w(2w)}|< t_{1s(2s)}$. Then we introduce a new parameter $\eta=t_{1w(2w)}/t_{1s(2s)}$, and obtain the full phase diagram for the modulated Haldane model (MHM) in the three-dimensional parameter-space, $(m, \phi, \eta)$, where $m$ is the inversion-breaking on-site potential term. We find that due to the breaking of the TR symmetry ($\phi\neq 0, \pm \pi$), the critical gap-closing states are semimetals with a single Dirac node. While $m$ is muted, phase transitions, accompanying the band-gap-closing-and-reopening, are observed at $\eta=\pm 0.5$ from a QAHI to an insulating phase with zero Chern number, which is demonstrated to be a HOTI (Fig.\ref{fig1}(b)) protected by inversion symmetry $\mathcal{I}$.


For any 2D insulator with zero Chern number ($C=0$), Wannier functions (WFs) of the occupied bands~\cite{Vanderbilt1997} can be used to identify topologically nontrivial states. Notably, the symmetric WFs associated with smooth and symmetric Bloch wave functions in momentum space may or may not be well defined~\cite{Vanderbilt2011}. 
In general, in a topologically nontrivial insulator, a set of symmetric WFs cannot be found for all its occupied bands~\cite{Po2017SI,Bernevig2017TQC}. Although symmetric WFs can be constructed, in the nontrivial state there is {\it mismatch} between the Wannier centers (WCs) and the lattice sites~\cite{FangPRL2017, EZAWA_1, EZAWA_2}. Wcs can be formulated in terms of $d$-dimensional polarization.\cite{Resta1992,Vanderbilt1993,Resta1994} The position of a WC is tied to a symmetric invariant point and the projection is quantized, and thus it can serve as a topological index for defining a topological insulator.

We will explore the evolution of topological states on various types of  edges within the framework of the MHM by examining the locations of WCs and their mismatch with the lattice sites. We then calculate the polarization $p_{\alpha=x,y}$~\cite{multiple}. A quantized nonzero polarization indicates that the insulator is topologically nontrivial. Finally, The HOTI phase is demonstrated in a properly designed nanodisk in which the in-gap zero-energy modes emerge. If one electron is filled in the zero mode, a $1/2$ fractional charge is distributed at each corner, featuring the topological nature of a 2D HOTI (Fig.~\ref{fig1}(b)). Interestingly, we find that the trivial insulator can transition to the HOTI without gap-closing along certain paths as  $\mathcal{I}$-symmetry is broken in the process.

{\it Modulated Haldane model and the phase diagram--}
Haldane model was proposed for the realization of the QAHI, or Chern insulator on a honeycomb lattice in the absence of an external magnetic field\cite{Haldane}. The essential ingredient of the model is the introduction of time-reversal breaking phase $\phi$ on the NNN hopping (see Fig.~\ref{fig1}(a)), which induces zero net magnetic flux per plaquette, so that the translational symmetry is preserved. The three-fold rotational symmetry $\mathcal{C}_3$ is preserved in Haldane model, that means, the hopping strength within NN and NNN are equal, {\it i.e.} $t_1 = t_{1s}=t_{1w}$ and $t_2 =t_{2s}=t_{2w}$ in Fig.\ref{fig1}(a). The Haldane model can be written as,

\begin{equation}
H = t_1\sum_{<i, j>} c_i^\dagger c_j + t_2 \sum_{\ll i, j\gg} e^{-iv_{ij} \phi}c_i^\dagger c_j + m\sum_{i}\epsilon_i c_i^\dagger c_i
\end{equation}
Here $v_{i, j} = sign(\hat{d_1} \times \hat{d_2})_z = \pm 1$ account for the alternating sign of magnetic phase, in which $\hat{d}_{1, 2}$ are the vectors along the two bonds constituting the next nearest neighbors. $m$ is the inversion-breaking on-site potential, with $\epsilon_i = \pm 1$ depending on whether $i$ is on the A or B sublattice. In Haldane model, the phase transition between a QAHI and a trivial insulator occurs when $m=\pm 3\sqrt{3}t_2 sin\phi$, and the gap-closing-and-reopening happens at the $\mathcal{C}_3$ symmetric points $K$ or $K'$.

In this work, we propose a modulated Haldane model (MHM) by making the hopping strength within the NN and NNN unequal as shown in Fig.~\ref{fig1}(a), where $|t_{1w(2w)}|< t_{1s(2s)}$. It is equivalent to applying a uniaxial strain along  ${x}$-direction to the honeycomb lattice. Consequently, $\mathcal{C}_3$ is broken. The Hamiltonian of the MHM takes the form $H(\mathbf{k}) = \sum_{i = 0} ^ {3} d_i(\mathbf{k}) \sigma_i$, where $\sigma_{i=0,1,2,3}$ are the identity matrix and three Pauli matrices, and
where 

\begin{eqnarray}
\label{H_MHM}
\nonumber
d_0 &=& 2 \text{cos}\phi[t_{2s} (\text{cos}\mathbf{k\cdot a_1} + \text{cos}\mathbf{k\cdot a_2})  + t_{2w} \text{cos}\mathbf{k}\cdot(\mathbf{a_1}-\mathbf{a_2})] ,\\ \nonumber
d_1 &=& t_{1s} \text{cos}\mathbf{k}\cdot\mathbf{\delta_1} + t_{1w} (\text{cos}\mathbf{k}\cdot\mathbf{\delta_2} + \text{cos}\mathbf{k}\cdot\mathbf{\delta_3}),\\ \nonumber
d_2 &=& t_{1s} \text{sin}\mathbf{k}\cdot\mathbf{\delta_1} + t_{1w} (\text{sin}\mathbf{k}\cdot\mathbf{\delta_2} + \text{sin}\mathbf{k}\cdot\mathbf{\delta_3}), \\ \nonumber
d_3 &=& 2\text{sin}\phi [t_{2s} (\text{sin}\mathbf{k}\cdot\mathbf{a_1} - \text{sin}\mathbf{k}\cdot\mathbf{a_2}) - t_{2w} \text{sin}\mathbf{k}\cdot(\mathbf{a_1}-\mathbf{a_2})] \\
     &&+ m . 
\end{eqnarray}

\begin{figure*}
\includegraphics[width=7 in]{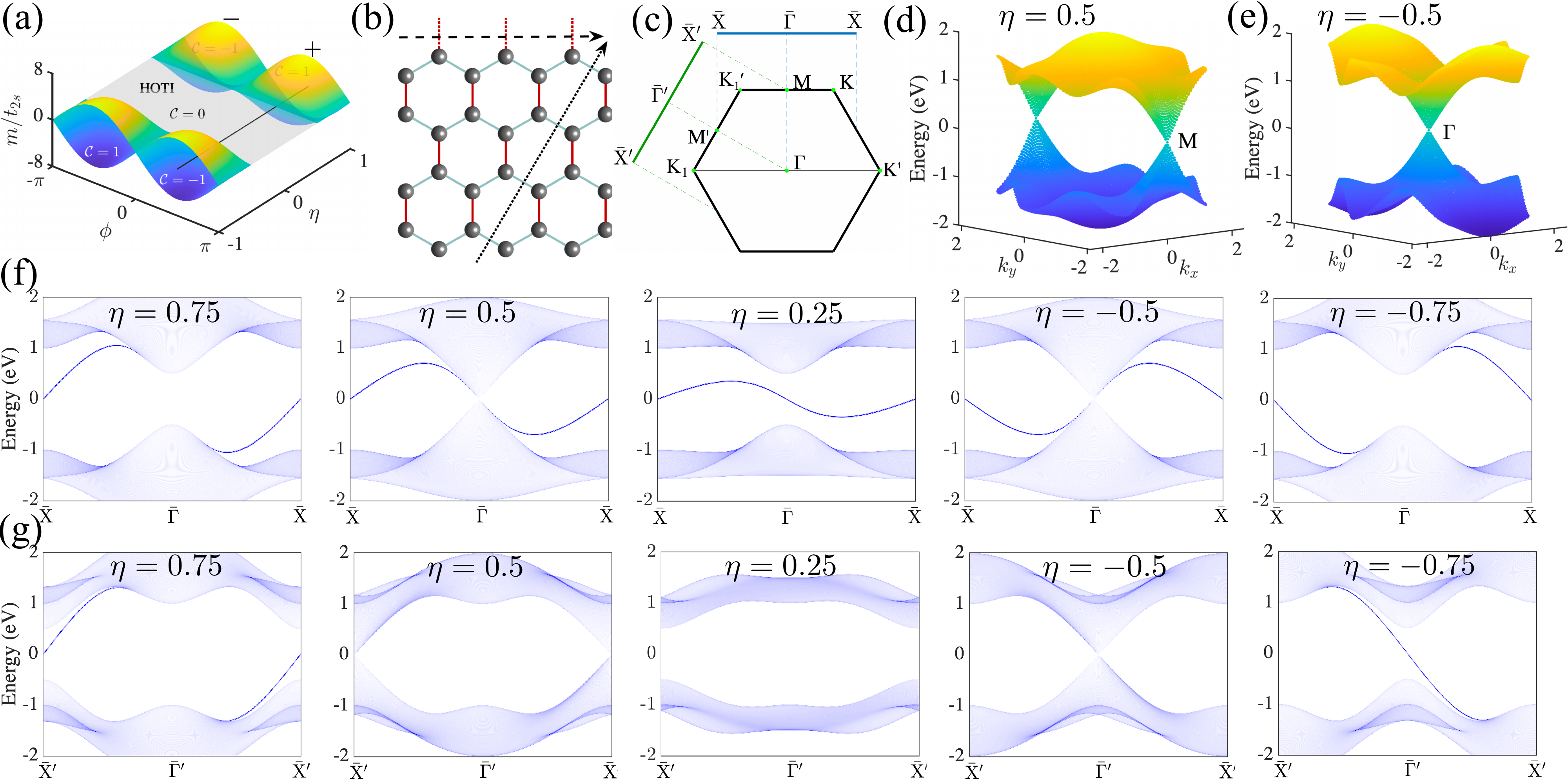}
\caption{(a) Phase diagram of the modulated Haldane model (MHM). (b) Honeycomb lattice and two types of zigzag edges labeled by the dashed and dotted lines, which cut strong (in red) and weak (in cyan) bonds, respectively. (c) The Brillouin zone and its projection on the two types of zigzag direction. (d, e) Band structures of critical semimetal phase which hosts one single Dirac node at (d) M point when $\eta=0.5$ and (e) $\Gamma$ point when $\eta=-0.5$, respectively. (f, g) The evolution of band structures for the two types of zigzag nanoribbon edges, (f) strong-bond-cut edge and (g) weak-bond-cut edge, as labeled in (b).}
\label{fig2}
\end{figure*} 

We can understand the behavior and topological nature of the MHM by looking at gap-closing-and-reopening transitions. Therefore, by introducing a new parameter $\eta= {t_{1w}}/{t_{1s}} ={t_{2w}}/{t_{2s}}$, we obtain a generalized phase diagram in 3D parameter space $(m, \phi, \eta)$ as shown in Fig.~\ref{fig2}(a), where $\phi\in [-\pi, \pi]$, and $\eta\in [-1,1]$. From Eq.~\ref{H_MHM}, we can find that the gap-closing phase transition happens at
\begin{equation}
m = \pm 3 t_{2s} \text{sin}\phi \sqrt{4-\frac{1}{\eta^2}}.
\end{equation}

In Fig.~\ref{fig2}(a), we can find the parameter space is divided into three sub-spaces by the pocket-like surfaces, on which the MHM falls in Dirac semimetal (DSM) phases and the band structures host one single Dirac node (DN) if $\mathcal{T}$ is broken with $\phi\neq 0, \pm\pi$. Otherwise, $\mathcal{T}$ symmetry enforces a pair of Dirac nodes. We find that the location of the Dirac node is $\eta$-dependent only. 
For each $\eta\in[-1,-0.5) \cup (0.5,1]$, there are two sinusoids, which we denote as $+$ and $-$, respectively. In regular Haldane model ($\eta = 1$), the DN was locked at the corner of Brillouin zone by the $\mathcal{C}_3$ symmetry. As $\eta$ varies from $1$ to $1/2$, the $\mathcal{C}_3$ is broken, the DN for $+$ ($-$) sinusoid no longer stays at $\mathcal{C}_3$ symmetric point  $K$ ($K'$), but moves from $K\rightarrow M$ ($K'_1 \rightarrow M$), and when $\eta <-1/2$, the DN moves along $\Gamma\rightarrow K'$ ($\Gamma\rightarrow K_1$)  (Fig.~\ref{fig2}(c)). When the two sinusoids meet at $\phi =0, \pm \pi$ where $\mathcal{T}$ survives, there exist two DNS inside the Brillouin zone related each other by $\mathcal{T}$. When $\eta = -1$, the two DNS are settled down at the midway of $\Gamma-K_1$ and $\Gamma-K'$, respectively. We find that the Dirac node is protected by the symmetry $\mathcal{M}_x*\mathcal{T}$, the combination of TR symmetry and the mirror symmetry normal to $x$-axis.

It seems that any two states in the out-of-pocket space can continuously deform to each other without gap-closing, so they may share the same topology. However, it turns out not to be true. Because the symmetries varies in the out-of-pocket parameter space, and it is possible that the phase transition occurs from state $\psi_A$ to $\psi_B$, without gap-closing while the symmetry changes in the path connecting $\psi_A$ and $\psi_B$. Therefore, we would like to figure out the symmetries preserved in each area, {\it i.e.}, the symmetry $\mathcal{C}_3$, $\mathcal{T}$, and $\mathcal{I}$ will survive at the plane $\eta=1$, $\phi=0,\pm\pi$ and $m=0$, respectively. For a generic point in the parameter space, $\mathcal{M}_x*\mathcal{T}$ is the only symmetry preserved. 

Naturally, we want to study the phases on these critical planes. We first will focus on the inversion symmetric plane with $m=0$.  Now we start from the QAHI ($\eta=1$) with a fixed $\phi=\pi/2$, and trace the phase transitions along the path labeled in Fig.~\ref{fig2}(a). As $\eta$ varies, one can observe the gap-closing-and-reopening at $\eta = \pm 1/2$, which indicates the phase transitions happen from a QAHI to an unknown insulating phase when $\eta\in (-1/2,1/2)$, whose Chern number is indicated as $C=0$. 

To better understand the phase transition, we make two types of zigzag edges to explore the evolution of the edge states in the phase transition, as shown in Fig.~\ref{fig2}(b), where the dashed (dotted) line cuts the strong (weak) hopping bonds, whose results are displayed in  Fig.~\ref{fig2}(f) (Fig.~\ref{fig2}(g)). In the calculation, we choose $t_{1s} = 1.0 eV$, $t_{2s} = 0.3 eV$. 
One can find that the MHM stays as a QAHI with Chern number $C = 1$ when $\eta \in (0.5, 1]$, and each edge supports one chiral gapless edge mode around $\bar{X}$(${\bar{X}}'$). While lowering $t_{1w,2w}$, at $\eta = 0.5$, the band-gap annihilates, leading to the emergence of one single Dirac node (\ref{fig2}(d)). As we mentioned above, this Dirac node is protected by $\mathcal{M}_x*\mathcal{T}$ and is pinned at M point due to the $\mathcal{I}$ symmetry, whereas the Dirac node is located at $K$ or $K'$ point in the regular Haldane model.
The Dirac point at $M$ point in MHM indicates a distinct band inversion, and implies that MHM may go to a nontrivial insulating phase, which is also suggested by the evolution of the edge states in Fig.~\ref{fig2}(f-g). Due to the distinct projecting locations of $M$ point on the two types of edges, {\it i.e.} the bulk Dirac node is projected at $\bar{\Gamma}$ (${\bar{X}}'$) for the dashed (dotted) line cut, their corresponding edge states are essentially different. For the weak-bonds-cut (dotted line), when the band-gap closes at ${\bar{X}}'$, the gapless edge mode is absorbed by bulk bands, and vanishes when the gap reopens. However, for the strong-bonds-cut-edge (dashed line), the edge mode survives. In contrast, the conduction and valence band-touching makes the edge mode connect to itself at $\bar{\Gamma}$, leading to the emergence of one in-gap edge mode. Unlike the topological edge mode of the QAHI, the edge mode here is detached from the bulk states, and is not topological. As $\eta$ changes the sign and, another band-closing-and-reopening  happens at $\Gamma$ point when $\eta=-0.5$, and the phase transits back to a Chern insulator, accompany with the emergence of the chiral gapless edge mode with opposite propagating direction, which indicates the Chern number changes to $C=-1$.

Comparing the two types of edges states for the insulator, the strong-bonds-cut supports a in-gap edge mode, but none for the weak-bonds-cut, which suggests that, in this insulating phase, the Wannier centers (WCs) may be located at the strong bonds, but not on the weak bonds. In other words, the in-gap edge mode can be expected for edges which cuts through the strong bonds, and it is demonstrated by edge states of the two types of armchair edges (see supplementary material). 


{\it Wannier Center and Higher-Order Topology--}
To verify the speculation and identify the topology of the insulator when $\eta \in (-0.5, 0.5)$, we calculate the WC associated with that phase on the $m=0$ plane. The WC, given by the polarization $(p_x, p_y)$, describes the average charge positions in a unit cell. The polarization can be written as 
\begin{equation}
p_{\alpha=x,y} = -\frac{1}{V}\int_{BZ} d\textbf{k} A_\alpha
\end{equation}
where $A_\alpha = -i<\psi|\partial_{k_\alpha}|\psi> $ is the Berry connection, $V$ is the volume of the Brillouin zone, and the integration is carried out over the whole Brillouin zone\cite{EZAWA_1, EZAWA_2, anomalous_hoti, multiple}. Due to the gauge invariance, the polarization is defined as $p_\alpha$ $\it{mod}$ $\mathbf{a_i}$, where $\mathbf{a_i}$ are lattice translation vectors. Here, we found the WC lies at the center of the strong bond. In general, since the WFs are symmetric, WCs should be located at the invariant point of the symmetry employed. On the $m=0$ plane, the inversion symmetry $\mathcal{I}$  is intact and ties the WC to the center of the strong bond, which is inversion invariant. The mismatch between the quantized WC and atomic site serves as topological index to characterize its higher-order topology.


Moreover, away from the inversion symmetric plane ($m\neq 0$), the WC will move toward the lattice site A (B) if $m > 0$ ($m<0$). When $|m|$ goes to infinity, the WC moves to the lattice site, and the system transits to an atomic insulator. In this case, the only symmetry involved is $\mathcal{M}_x*\mathcal{T}$, under which $p_y$ cannot be quantized. In other words, while $\mathcal{I}$ symmetry is broken, a phase transition occurs, from a HOTI to a trivial insulator, without gap-closing.

\begin{figure}
\begin{centering}
\includegraphics[width=\linewidth]{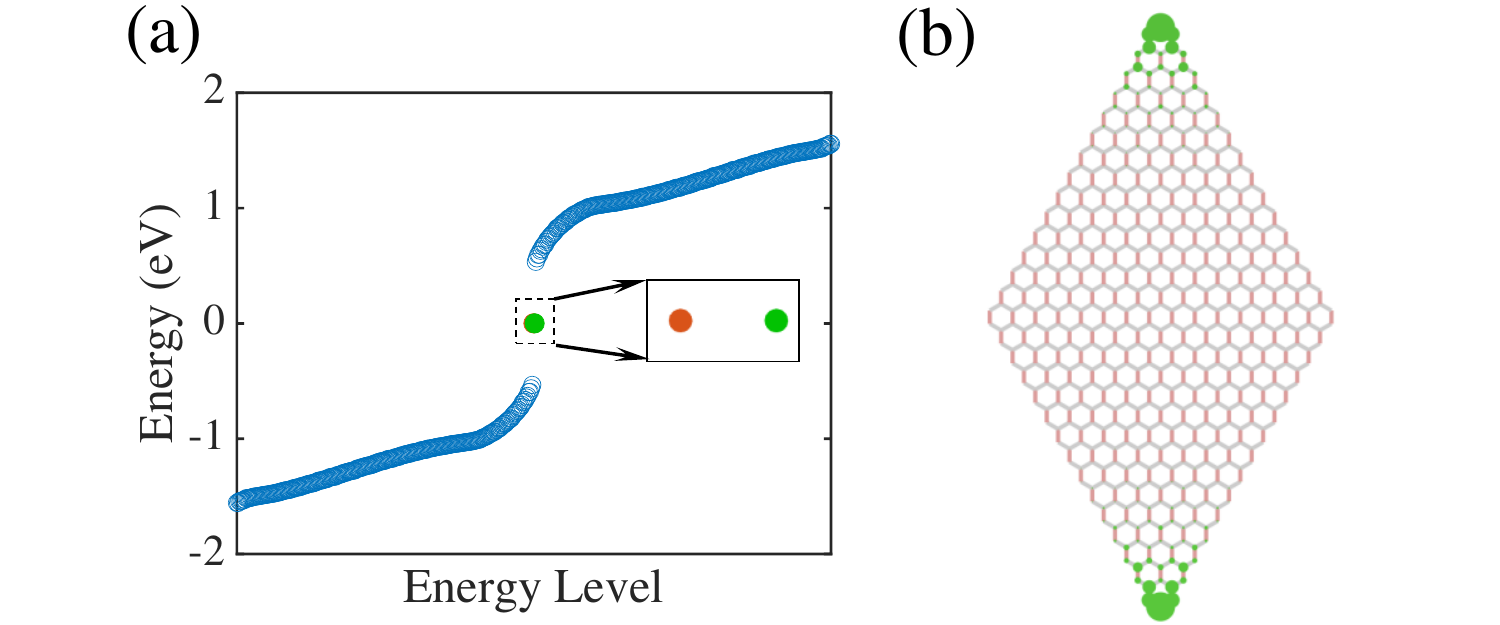}
\par\end{centering}
\centering{}
\caption{(a). The calculated energy spectrum on the nanodisk when the system is in the HOTI phase. There are two in-gap zero-energy modes as shown in the inset. (b). The charge distribution of the zero-energy mode indicates that if one mode (marked in green) is filled with one electron, a fractional $1/2$ will be distributed at each of the two corners.}
\label{fig3}
\end{figure}

A HOTI features the emergence of 0D zero-energy corner mode in a properly designed nanodisk, which is demonstrated for our case (Fig.~\ref{fig3}). 
We calculate the eigenstates of the nanodisk. Fig.~\ref{fig3}(a) plots the energy levels. There are two degenerate in-gap zero-energy modes, marked in green and orange, respectively. If one mode is filled with an electron, a $1/2$ fractional charge will be distributed at each of the two corners, as illustrated in Fig.~\ref{fig3}(b). As long as the inversion symmetry breaks, the in-gap zero-modes will disappear, and the HOTI phase will become trivial.

{\it Conclusion.--}
We explore topological properties of a modulated Haldane model (MHM) in which we have made the hopping strength unequal between the NN and NNN terms and broken the three-fold rotational symmetry $\mathcal{C}_3$ of the Hamiltonian. MHM is found to harbor a new nontrivial insulating phase on the inversion-symmetric plane, which can transition to the trivial insulator state without gap-closing. The HOTI nature of the nontrivial phase is demonstrated by showing that the quantized Wannier centers, which are enforced by the inversion symmetry, do not lie at the lattice sites. The 2D HOTI state features a zero-dimensional, zero-energy corner mode, which is confirmed in the spectrum of a properly designed nanodisk. 
Interestingly, we find that the HOTI phase can transition to a trivial insulator without gap-closing along certain paths in the phase diagram on which the system undergoes a change in the inversion symmetry. The gap-closing critical states are magnetic semimetals with a single Dirac node, which is not constrained to lie at the high-symmetry points in the Brillouin zone as is the case for the Haldane model. Our modulated Haldane model provides a rich playground for exploring HOTI and other topological and non-topological phases and how transitions can be induced between these phases by varying the parameters in the model.

{\it Acknowledgement. } This work is supported by the Air Force Office of Scientific Research under award number FA955-20-1-0322, and it benefited from computational resources of Northeastern University's Advanced Scientific Computation Center (ASCC) and the Discovery Cluster.

\bibliographystyle{apsrev4-1}

%

%

\clearpage
\widetext
\setcounter{equation}{0}
\setcounter{figure}{0}
\setcounter{table}{0}
\renewcommand{\thepage}{S\arabic{page}}
\renewcommand{\thesection}{S\arabic{section}}
\renewcommand{\theequation}{S\arabic{equation}}
\renewcommand{\thefigure}{S\arabic{figure}}
\renewcommand{\thetable}{S\arabic{table}}
\renewcommand{\bibnumfmt}[1]{[S#1]}
\renewcommand{\citenumfont}[1]{S#1}
\newcommand{\bk}{\boldsymbol\kappa}

\newcommand{\beginsupplement}{%
  \setcounter{equation}{0}
  \renewcommand{\theequation}{S\arabic{equation}}%
  \setcounter{table}{0}
  \renewcommand{\thetable}{S\arabic{table}}%
  \setcounter{figure}{0}
  \renewcommand{\thefigure}{S\arabic{figure}}%
  \setcounter{section}{0}
  \renewcommand{\thesection}{S\Roman{section}}%
  \setcounter{subsection}{0}
  \renewcommand{\thesubsection}{S\Roman{section}.\Alph{subsection}}%
}

\section{Supplemental Material}

\begin{figure*}
\includegraphics[width=3 in]{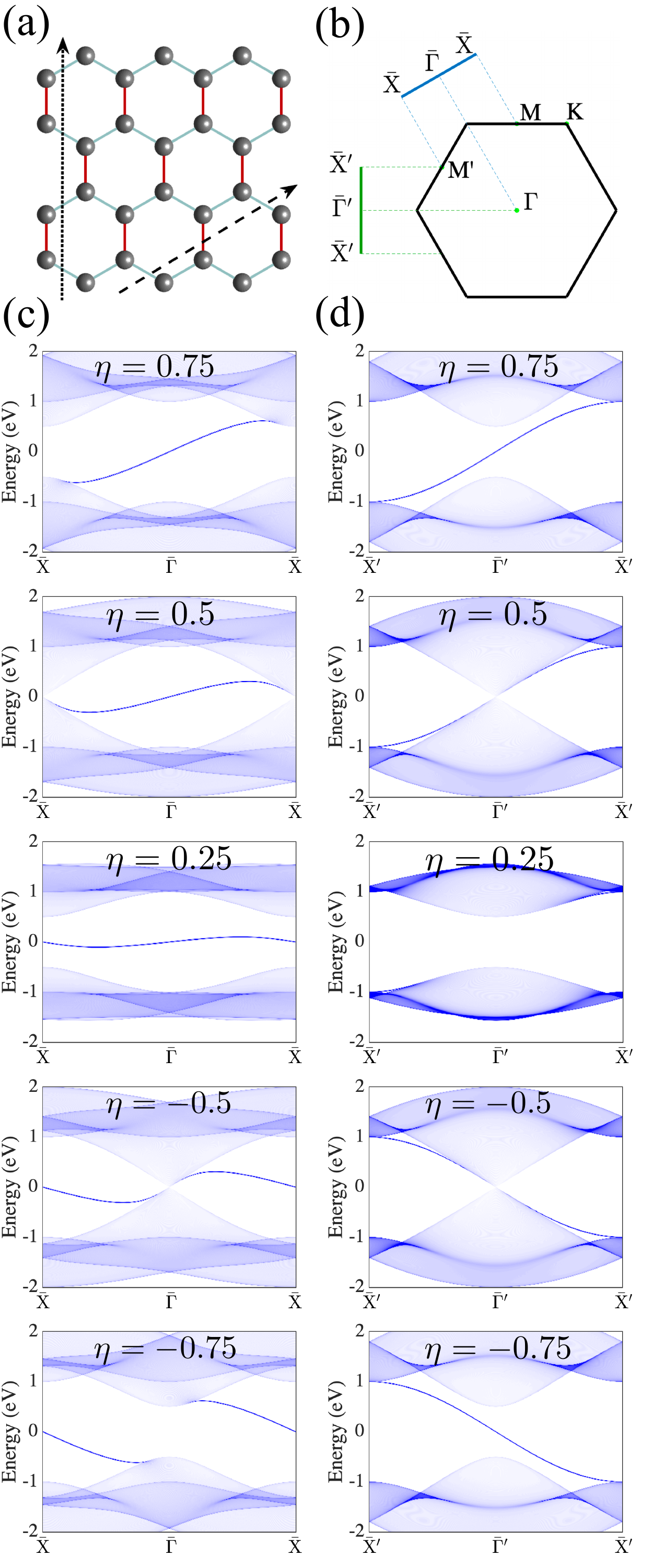}
\caption{(a) The honeycomb lattice. Dashed and dotted arrows label two types of armchair nanoribbon edges, which cut strong (marked in red) and weak (marked in light cyan) bonds, respectively. (c) The Brillouin zone and its projection on the two types of armchair direction. (c, d) The evolution of edge band structures for the two types of armchair nanoribbon, strong-bond-cut (c) edge and weak-bond-cut edge(d), as labelled in (a).}
\label{fig:s1}
\end{figure*}

Here, we explore the evolution of band structures of two types of armchair nanoribbons as illustrated in Fig.~\ref{fig:s1}(a, b). The dashed arrow labelled armchair edge cuts both the strong and weak bonds, while the dotted arrow labelled edge cuts only the weak bonds. We use the same set of parameters as in the main text, $i. e.  t_{1s} = 1.0 eV, t_{2s} = 0.3 eV, m = 0$ and $\phi = \pi/2$. When $\eta\in(0.5, 1]$, the modulated Haldane model (MHM) supports a Chern insulator phase with $\mathcal{C}=1$, therefore, one gapless edge band appears in the band gap, which applies both to the two types of edges (Fig.~\ref{fig:s1}(c, d)). At $\eta = 0.5$, the bulk band gap closes at $M$ (see Fig. 2(d) in main text). Accordingly, the edge band spectrum closes the gap at $\bar{X}$ and $\bar{\Gamma}'$ for the two types of edges. For $\eta\in(-0.5, 0.5)$, the Haldane model is an insulating phase with $\mathcal{C}=0$. For the dashed line labelled edge, an edge band emerges inside the gap and is detached from the bulk bands (Fig.~\ref{fig:s1}(c)). While for the dotted line labelled edge, the edge band is absorbed into the bulk spectrum and vanishes (Fig.~\ref{fig:s1}(d)). It indicates that the Wannier center is located on the strong bond, which agrees with the result from zigzag nanoribbon as discussed in the main text. Another band gap closing occurs at $\eta = -0.5$, where the gap closes at $\Gamma$ (see Fig. 2(e) in main text). When $\eta\in[-1, -0.5)$, the modulated Haldane model transits back to Chern insulator but carries a different number $\mathcal{C}=-1$. 

\end{document}